\documentclass[twocolumn,english,aps,pra,superscriptaddress,showpacs,tightenlines]{revtex4}
\usepackage{amsmath}
\usepackage{amssymb}
\usepackage{graphicx}

\begin{document}


\title{Non-Abelian geometrical control of a qubit in an NV center in diamond}

\author{Jing \surname{Lu}}
\affiliation{Key Laboratory of Low-Dimensional Quantum Structures and Quantum
Control of Ministry of Education, and Department of Physics, Hunan
Normal University, Changsha 410081, China}
\author{Lan \surname{Zhou}}
\email{zzhoulan@gmail.com}
\affiliation{Key Laboratory of Low-Dimensional Quantum Structures and Quantum
Control of Ministry of Education, and Department of Physics, Hunan
Normal University, Changsha 410081, China}

\begin{abstract}
We propose an approach for an optical qubit rotation in the negatively charged
nitrogen-vacancy (NV) center in diamond. The qubit is encoded in the ground
degenerate states at the relatively low temperature limit. The basic idea of
the rotation procedure is the non-Abelian geometric phase in an adiabatic passage,
which is produced by the nonadiabatic transition between the two degenerate dark
states. The feasibility is based on the success of modeling the NV center as an
excited-doublet four-level atom.
\end{abstract}

\pacs{42.50.Ex, 61.72.-y, 03.65.Vf}
\maketitle

\section{\label{Sec:1}Introduction}

Phases play a major role in all interference and diffraction phenomena in
optics, wave physics and quantum mechanics. The study of phase has generated
a large number of practical device~\cite{Bpr281}: superconducting
interference device, interferometers to measure lengths and small rotations,
modulators to modulate wavefront, etc. In theoretical physics, phases are
involved in the distinction between fermions and bosons, decoherence of a
quantum system and Aharonov-Bohm effect. It is well known that a quantum
system acquire a Berry phase~\cite{Berry84} which evolves adiabatically
around a circuit by varying parameters in its Hamiltonian. This phase is a
sum of two parts. The first one is related to the instantaneous energy of
the system, which is called dynamical phase. The second one is a purely
geometric property of the closed path followed by the system in space, which
is called geometrical phase. Later, non-Abelian gauge fields arising in the
adiabatic development of simple quantum mechanical systems are introduced by
Wilczek \textit{et al}~\cite{Wilczek}. Nowadays, the Berry phase is not only
of interest from a fundamental point of view, for example, the induced gauge
field~\cite{RMP01523,sunPRD} and topological states of matter~\cite{graphene}%
, but also may have important applications in quantum information
processing. Holonomic quantum computation~\cite{HQC} is a prototype of
research on quantum gates based on Abelian or non-Abelian geometric phases.
Its significant attraction lies on that these quantum gates could be
inherently robust against some local perturbations. Moreover, interest are
paid on the adiabatic evolution of dark eigenstates of the considered system
in order to remove any accompanied dynamical phase shift.

On the other hand, the negatively charged nitrogen-vacancy (NV) center in
diamond has recently emerged as a promising candidate for practical and
scalable implementation of quantum information processing due to the
following robustness: it is an individually addressable quantum system; its
quantum state can be initialized and manipulated with optical and microwave
fields, and even measured with high fidelity at room temperature\cite%
{JPCM18,Lukins316,greent06,stoneham21,PNAS107}. NV centers are formed by a
substitutional nitrogen atom adjoining to a vacancy in the diamond lattice.
The dangling bonds near the vacancy are occupied by six electrons, which
includes three dangling bonds on the carbon atoms and two bonding bonds on
the nitrogen atom~\cite{Lenef96}. The nitrogen atom breaks down the $T_{d}$
symmetry and gives rise to the trigonal symmetry~\cite{Hollen13,LukinNJP13}.
A single NV center have a long-lived spin triplet in its electronic ground
state whose levels function as a qubit and are tunable with an applied
magnetic field~\cite{Jelezko92}. Currently, the applications rely on the
spin in the ground state, for example, few-qubit networks for simple
algorithms and quantum memories are created by coherently coupling NV-center
spins to nearby electronic~\cite{Neumann6,DuPRL105} and nuclear spins~\cite%
{Lukins316,Fuchs11}; the spin triplet ground state also acts as a sensitive
magnetic probe of the local environment~\cite{Hanson320,Zhaonat6,Zhao106}.
Optical techniques constitute powerful tools in quantum physics and quantum
information science~\cite{NP5(11)335,NP4(10)211}. It is known for a
considerable time that the NV in its singly-charged state is good
single-photon source~\cite{NVSphotons}. Nowadays, its optical Rabi
oscillations have been observed~\cite{batalov100}. Coherent population
trapping using optical laser fields has been found in single NV centers~\cite%
{cptNV97}. And the crucial element for solid-state realization of quantum
optical networks has been demonstrated by the entanglement between the
polarization of a single optical photon and a solid-state qubit associated
with the single electronic spin of a NV center in diamond~\cite{Togan466}.

An important advantage of optical techniques in comparison with microwave
manipulation is spatial resolution which can selectively address a single NV
defect in diamond as well as individual qubits. Here, we propose an scheme
to achieve an qubit rotation by the coherent interaction of light and single
NV centers in diamond. The qubit is encoded in the degenerate Zeeman
sublevels in the absence of external strain and electric or magnetic fields.
Due to the spin--orbit and spin--spin interactions, an excited-double
four-level system can be established in the spin-triplet ground state and an
orbital-doublet, spin-triplet excited state of the NV center, which allow us
to apply two pump and two Stokes laser pulses. The rotation of qubit based
on the non-Abelian geometric phase is realized by tuning adiabatically the
Rabi frequencies and phases of the pump and Stokes pulses.

The paper is organized as follows. In Sec.~\ref{Sec:2}, the energy levels of
a NV center is briefly reviewed and the excited-double four-level system is
modeled for a NV center with an electronic spin triplet in the ground state
and an orbital-doublet, spin-triplet excited state. In Sec.~\ref{Sec:3}, we
present our method for arbitrary qubit rotation. Summaries are made in Sec.~%
\ref{Sec:4}.

\section{\label{Sec:2} Formulation of an excited-double four-level system}

NV centers consist of a substitutional nitrogen atom and an adjacent
vacancy, which are a naturally occurring defect in diamond with highly
localized electronic bound states. The bound states of NV centers are
multiparticle states of six electrons: five contributed by the
nearest-neighbor nitrogen and carbon atoms to the vacancy and an additional
captured from the bulk. To find out the electronic states of NV centers,
single-electron molecular orbitals are employed, which are built up from
linear combinations of the dangling sp3 orbitals of one nitrogen and three
carbon atoms around the vacancy. Applying group theoretical arguments to the
NV center leads to four molecular orbitals $\left\{ u,v,e_{x},e_{y}\right\} $%
\cite{Lenef96,Hollen13,LukinNJP13}. Orbitals $u$ and $v$ are totally
symmetric but non-degenerate which transform according to the
one-dimensional irreducible representation of $C_{3v}$ symmetry, and the two
degenerate orbitals $\left\{ e_{x},e_{y}\right\} $ transform according to
the two-dimensional irreducible representation of $C_{3v}$ point group. The
ordering of the orbitals are obtained by symmetry and charge considerations
of the electron-ion interaction, i.e. $u$ orbital has the lowest energy, $v$
is the next lowest orbital and doubly degenerate $e_{x,y}$ orbitals are the
highest energy. Filling four orbitals with six electrons according to the
Pauli exclusion principle indicates that the ground state consists of four
electrons completely occupying the totally symmetric orbitals and two
electrons occupying the remaining orbitals. Although the spin-obit wave
functions for the ground-state configuration give $^{3}A_{2}$, $^{1}A_{1}$
and $^{1}E$ states due to the antisymmetry property of the total
wavefunction for fermionic particles, Hund's rules predict that the ground
state is attributed to the spin-triplet $^{3}A_{2}$ state. The excited state
configuration can be learned from one electron being promoted from the $v$
orbital to the $e_{x,y}$ orbitals in the single particle picture~\cite%
{PNAS107}. The excited-state configuration has an orbital-doublet,
spin-singlet $^{1}E$ state and an orbital-doublet, spin-triplet $^{3}E$
state with manifold $\left\{ E_{x},E_{y},E_{x}^{\prime },E_{y}^{\prime
},A_{1},A_{2}\right\} $ according to the irreducible representation of $%
C_{3v}$ symmetry. The degeneracy between triplets and singlets in the
ground-state and excited-state configurations are lifted by the
electron--electron Coulomb interaction. Nowadays, it is known that the
optical transition is associated with the $^{3}A_{2}$ ground state and $%
^{3}E $ excited state. The spin--spin interaction causes the 2.87GHz zero
field splitting between the magnetic sublevels $m_{S}=0$ and $m_{S}=\pm 1$
states of the $^{3}A_{2}$ ground triplet state. The spin--orbit interaction
splits the $^{3}E$ excited state into three doublet degenerate states, noted
by $E,E^{\prime },A_{1}$ and $A_{2}$. And the degeneracy of the $A_{1}$ and $%
A_{2}$ states is lifted by the spin--spin interaction.
\begin{figure}[tbp]
\includegraphics[bb=65 515 290 803, width=3 cm]{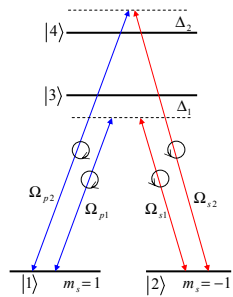}
\caption{(Color online) The energy diagram and selection rule of the
excited-doublet four-level system associated with the ground and excited
states in the negatively charged nitrogen-vacancy center.}
\label{fig2:1}
\end{figure}

The $\sigma _{+}\left( \sigma _{-}\right) $-polarized optical pulse drives
the transition between the ground $m_{s}=-1\left( +1\right) $ state with the
$A_{1}$ and $A_{2}$ states. Neglecting the $\sigma _{+}\left( \sigma
_{-}\right) $-polarized optical coupling between the ground $m_{s}=1\left(
-1\right) $ state and the $E_{x}$ and $E_{y}$ states and the $x\left(
y\right) $-polarized optical coupling between the ground $m_{s}=0$ state and
the $E_{y}^{\prime }\left( E_{x}^{\prime }\right) $, a excited-doublet
four-level system is therefore configured in the absence of external strain
and electric or magnetic fields, whose energy diagram and selection rule are
schematically illustrated in Fig.~\ref{fig2:1}. The ground states with $%
m_{s}=\pm 1$ are denoted by $\left\vert 1\right\rangle $ and $\left\vert
2\right\rangle $, and the excited states are denoted by $\left\vert
3\right\rangle $ and $\left\vert 4\right\rangle $. The transitions from
states $\left\vert 1\right\rangle $ with energy $\omega _{1}$ and $%
\left\vert 2\right\rangle $ with energy $\omega _{2}$ to the upper two
states with energy $\omega _{3}$ and $\omega _{4}$ are driven by the pump
fields and the Stokes laser fields, respectively. The pump field is a $%
\sigma _{-}$-polarized bichromatic wave with frequencies $\nu _{p1}$ and $%
\nu _{p2}$. And the Stokes field is a $\sigma _{+}$-polarized bichromatic
wave with frequencies $\nu _{s1}$ and $\nu _{s2}$. The Hamiltonian for this
system is then%
\begin{align}
H& =\omega _{1}\left\vert 1\right\rangle \left\langle 1\right\vert +\omega
_{2}\left\vert 2\right\rangle \left\langle 2\right\vert +\omega
_{3}\left\vert 3\right\rangle \left\langle 3\right\vert +\omega
_{4}\left\vert 4\right\rangle \left\langle 4\right\vert \\
& +\frac{i}{2}\left( \Omega _{p1}e^{-i\nu _{p1}t}-\Omega _{p2}e^{-i\nu
_{p2}t}\right) \left( \left\vert 1\right\rangle \left\langle 3\right\vert
-\left\vert 1\right\rangle \left\langle 4\right\vert \right) +h.c.  \notag \\
& -\frac{i}{2}\left( \Omega _{s1}e^{-i\nu _{s1}t}+\Omega _{s2}e^{-i\nu
_{s2}t}\right) \left( \left\vert 2\right\rangle \left\langle 3\right\vert
+\left\vert 2\right\rangle \left\langle 4\right\vert \right) +h.c.  \notag
\end{align}%
where $\Omega _{pi}$ and $\Omega _{si}$ are time-dependent Rabi frequencies
where a $\pi $ phase difference has been introduced in the pump field. By
assuming that frequencies $\nu _{pi}=\nu _{si}$, we can define two-photon
detunings $\Delta _{1}=\omega _{3}-\omega _{1}-\nu _{p1}$ and $\Delta
_{2}=\nu _{p2}+\omega _{1}-\omega _{4}$ due to the degeneracy of the ground
state. In the rotating frame with respect to
\begin{equation}
H_{0}=\sum_{i=1}^{2}\omega _{i}\left\vert i\right\rangle \left\langle
i\right\vert +\left( \omega _{3}-\Delta _{1}\right) \left\vert
3\right\rangle \left\langle 3\right\vert +\left( \omega _{4}+\Delta
_{2}\right) \left\vert 4\right\rangle \left\langle 4\right\vert \text{,}
\label{eq2:2}
\end{equation}%
the Hamiltonian in the interaction picture is obtained as%
\begin{equation}
H_{I}=\left[
\begin{array}{cccc}
0 & 0 & \frac{i}{2}\Omega _{p1} & \frac{i}{2}\Omega _{p2} \\
0 & 0 & -\frac{i}{2}\Omega _{s1} & -\frac{i}{2}\Omega _{s2} \\
-\frac{i}{2}\Omega _{p1}^{\ast } & \frac{i}{2}\Omega _{s1}^{\ast } & \Delta
_{1} & 0 \\
-\frac{i}{2}\Omega _{p2}^{\ast } & \frac{i}{2}\Omega _{s2}^{\ast } & 0 &
-\Delta _{2}%
\end{array}%
\right]  \label{eq2:3}
\end{equation}%
by employing the rotating-wave approximation.
\begin{figure}[tbp]
\includegraphics[width=8cm]{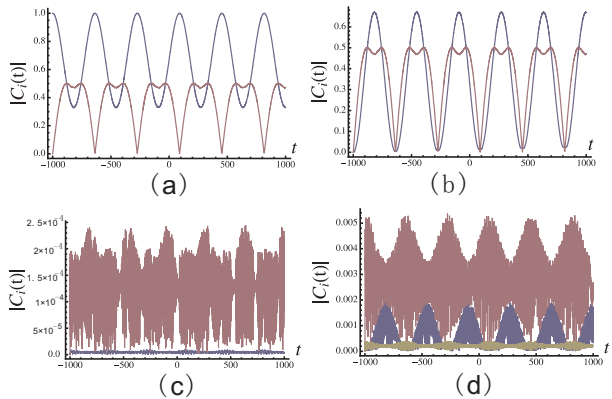}
\caption{(Color online) The norm of probability amplitudes for the $^3A_2$
ground and $^3E$ excited states as function of time when NV center is
initially in the $m_S=-1$ state of the $^3A_2$ ground triplet state, (a) for
$m_S=-1$ ground (blue) and $A_2$ excited (red) states, (b) $m_S=1$ ground
(blue) and $A_1$ excited (red) states. The norm of the amplitude for $m_S=0$
ground state is plotted in (c) with blue. The superpositions of states $%
E^{\prime }_x$, $E_x$, $E_y$ and $E^{\prime }_y$ give rise to other four
states whose norms are shown in (c) with red color and (d). All Rabi
frequencies are taken as $14$MHz. And detunings $\Delta _{1}=\Delta
_{2}=10MHz$. Other spin-orbit and spin-spin parameters are taken from Ref.~%
\protect\cite{Hollen13}. The time axis is in units of ns. }
\label{fig2:2}
\end{figure}

To justify that it is valid to neglect optical coupling between the ground $%
m_{s}=1\left( -1\right) $ state and the $E_{x}$ and $E_{y}$ states, as well
as the coupling between the ground $m_{s}=0$ state and the $E_{y}^{\prime
}\left( E_{x}^{\prime }\right) $, we take the spin-orbit and spin-spin
parameters from Ref.~\cite{Hollen13}. Besides, we set all Rabi frequencies $%
\Omega _{pi}$ and $\Omega _{si}$ have the same value $14MHz$, and detunings $%
\Delta _{1}=\Delta _{2}=10$MHz. Then we numerically solve the time-dependent
Schr\"{o}dinger equation for the spin-triplet $^{3}A_{2}$ state and
orbital-doublet, spin-triplet $^{3}E$ state with the system initially in the
$m_{S}=-1$ state of the $^{3}A_{2}$ ground triplet state. Figure~\ref{fig2:2}
numerically illustrates the norm of probability amplitudes as a function of
time for (a) $m_{S}=-1$ ground (blue) and $A_{2}$ excited (red) states, (b) $%
m_{S}=1$ ground (blue) and $A_{1}$ excited (red) states. The norm of the
amplitude for $m_{S}=0$ ground state is plotted in Fig.(c) with blue. The
superpositions of states $E_{x}^{\prime }$, $E_{x}$, $E_{y}$ and $%
E_{y}^{\prime }$ give rise to other four states whose norms are shown in
Fig.~\ref{fig2:2}(c) with red and (d). It can be found that the
probabilities on the ground $m_{S}=\pm 1$ states and $A_{1},A_{2}$ excited
states are much larger than those of other states. Therefore, we can
eliminate the contribution of the other states and model the NV centers as
excited-doublet four-level systems.

\section{\label{Sec:3}qubit rotation}

We now find the eigenstates of Hamiltonian~(\ref{eq2:3}). The dark states
would be desirable for gate operation in order to remove the dynamic phase
shift. Dark states are zero-eigenvalue eigenstates of a considered system.
The characteristic equation of this excited-doublet four-level system shows
that two eigenvalues of Eq.(\ref{eq2:3}) will be zero if we let frequency
detunings $\Delta _{1}=\Delta _{2}=\omega _{0}$, $\Omega _{pi}=\Omega _{p}$
and $\Omega _{si}=\Omega _{s}$. And these two corresponding dark states $%
\left\vert D_{1}\right\rangle $ and $\left\vert D_{2}\right\rangle $ are
coherent superpositions of the bare states and have the following expression
\begin{subequations}
\label{eq2:4}
\begin{align}
\left\vert D_{1}\right\rangle & =\frac{\Omega _{s}^{\ast }}{\Omega }%
\left\vert 1\right\rangle +\frac{\Omega _{p}^{\ast }}{\Omega }\left\vert
2\right\rangle , \\
\left\vert D_{2}\right\rangle & =\frac{i\Omega }{\sqrt{2}\Theta }\left(
\left\vert 3\right\rangle -\left\vert 4\right\rangle \right) +\sqrt{2}\frac{%
\omega _{0}}{\Theta }\left\vert B_{1}\right\rangle ,
\end{align}%
where $\Omega =\sqrt{\left\vert \Omega _{s}\right\vert ^{2}+\left\vert
\Omega _{p}\right\vert ^{2}}$ and $\Theta =\sqrt{\Omega ^{2}+2\omega _{0}^{2}%
}$. Here, the bright state
\end{subequations}
\begin{equation}
\left\vert B_{1}\right\rangle =\frac{\Omega _{p}}{\Omega }\left\vert
1\right\rangle -\frac{\Omega _{s}}{\Omega }\left\vert 2\right\rangle
\end{equation}%
is orthogonal to the dark state $\left\vert D_{1}\right\rangle $.

Let us first consider that the applied pump and Stokes fields are in
resonance with its corresponding atomic transition, and two-photon resonance
is still maintained. In this case, we have $\omega _{0}=0$. The dark state $%
\left\vert D_{2}\right\rangle $ is coherent superpositions of the bare
states $\left\vert 3\right\rangle $ and $\left\vert 4\right\rangle $, which
leads to $\left\langle D_{i}\right\vert \partial _{t}\left\vert
D_{j}\right\rangle =0$ for $i\neq j$, i.e. the non-adiabatic transition
between the two degenerate dark states vanishes. Then, one can completely
transfer the population from $\left\vert 1\right\rangle $ to $\left\vert
2\right\rangle $ vice versa by the techniques of Stimulated Raman adiabatic
passage since the pump and Stokes field are involved in the dark state $%
|D_{1}\rangle $.

When the applied fields are off resonance, i.e., $\omega _{0}\neq 0$, there
is a non-adiabatic transition between these two dark states, which plays the
important role in single-qubit gate operation. Notice that states $|1\rangle
$ and $|2\rangle $ span the same eigenspace as the dark state $\left\vert
D_{1}\right\rangle $ and bright state $\left\vert B_{1}\right\rangle $, and
the qubit described by $\alpha |1\rangle +\beta |2\rangle $ does not change
with time when $\Omega _{s}=\Omega _{p}=0$, we can project the qubit on the
dark state $\left\vert D_{1}\right\rangle $ and bright state $\left\vert
B_{1}\right\rangle $. The projection requires the Rabi frequencies $\Omega
_{p}$ and $\Omega _{s}$ to have essentially the same envelopes
\begin{subequations}
\label{eq2:5}
\begin{align}
\Omega _{p}\left( t\right) & =\Omega \left( t\right) \cos \chi e^{i\phi
}e^{i\psi \left( t\right) }, \\
\Omega _{s}\left( t\right) & =\Omega \left( t\right) \sin \chi e^{-i\phi
}e^{i\psi \left( t\right) },
\end{align}%
Here, $\chi $ and $\phi $ are time-independent angles, which specify the
Euler angle of a vector
\end{subequations}
\begin{equation*}
\left\vert \mathbf{D}\right\rangle =\sin \chi e^{-i\phi }\left\vert
1\right\rangle +\cos \chi e^{i\phi }\left\vert 2\right\rangle
\end{equation*}
The first dark state and bright state can be written as
\begin{equation}
\left\vert D_{1}\right\rangle =e^{-i\psi \left( t\right) }\left\vert \mathbf{%
D}\right\rangle ,\left\vert B_{1}\right\rangle =e^{i\psi \left( t\right)
}\left\vert -\mathbf{D}\right\rangle .  \label{eq2:6}
\end{equation}%
We denote the Bloch sphere containing vectors $\left\vert \mathbf{D}%
\right\rangle $ and $\left\vert -\mathbf{D}\right\rangle $ as BS1. The fixed
angles $\chi $ and $\phi $ keep vectors $\left\vert \mathbf{D}\right\rangle $
and $\left\vert -\mathbf{D}\right\rangle $ as constant vectors in BS1 during
the time evolution. Hence, any initial state of the qubit can be written as
a superposition of $\left\vert \mathbf{D}\right\rangle $ and $\left\vert -%
\mathbf{D}\right\rangle $. The mutually orthogonal state vectors $\left\vert
-\mathbf{D}\right\rangle $ and $\left\vert \mathbf{0}\right\rangle \equiv
(\left\vert 3\right\rangle -\left\vert 4\right\rangle )/\sqrt{2}$ construct
the north and south pole of another unit 2-sphere denoted as BS2. The second
dark state $\left\vert D_{2}\right\rangle $ is a vector in BS2, whose
adiabatic evolution can be written as%
\begin{equation}
\left\vert D_{2}\right\rangle =i\sin \varphi \left( t\right) \left\vert
\mathbf{0}\right\rangle +e^{i\psi \left( t\right) }\cos \varphi \left(
t\right) \left\vert -\mathbf{D}\right\rangle ,  \label{eq2:7}
\end{equation}%
where $\tan \varphi =\Omega /\left( \sqrt{2}\omega _{0}\right) $. Since the
two photon detuning is always on, the adiabatic pulses force the system to
remain in dark states. However, vector $\left\vert D_{2}\right\rangle $
moves slowly with time due to its time-dependent azimuthal and pole angles.
When the pump and Stokes pulses are adiabatically turned on, the eigenstate $%
\left\vert D_{2}\right\rangle $ evolves smoothly from the initial state $%
\left\vert -\mathbf{D}\right\rangle $ to a intermediate state which is a
superposition state of $\left\vert \mathbf{0}\right\rangle $ and $\left\vert
-\mathbf{D}\right\rangle $. The inverse process, i.e. adiabatically turning
of the pump and Stokes pulses, brings the system back to its initial state $%
\left\vert -\mathbf{D}\right\rangle $. When angles $\varphi $ and $\psi $
make a cyclic evolution, the path of the $\left\vert D_{2}\right\rangle $
vector encloses a patch on the BS2 with a solid angle $\gamma _{c}$, which
is the Berry phase. This cyclic evolution gives rise to a phase change on
the state $\left\vert -\mathbf{D}\right\rangle $. The unitary operator%
\begin{equation}
U_{C}=e^{i\gamma _{C}}\left\vert -\mathbf{D}\right\rangle \left\langle -%
\mathbf{D}\right\vert +\left\vert \mathbf{D}\right\rangle \left\langle
\mathbf{D}\right\vert   \label{eq2:8}
\end{equation}%
depicts the effect of the cyclic evolution on the vectors in BS1. In the
bare basis $\left\vert 1\right\rangle $ and $\left\vert 2\right\rangle $,
the unitary operator
\begin{equation}
U_{C}=e^{i\gamma _{c}/2}\left[ \cos \frac{\gamma _{c}}{2}\hat{I}+i\sin \frac{%
\gamma _{c}}{2}\hat{n}\cdot \hat{\sigma}\right]   \label{eq2:9}
\end{equation}%
rotates the qubit encoded in states $\left\vert 1\right\rangle $ and $%
\left\vert 2\right\rangle $ about the axe
\begin{equation}
\hat{n}=\left( -\sin 2\chi \cos 2\phi ,-\sin 2\chi \sin 2\phi ,\cos 2\chi
\right)   \label{eq2:10}
\end{equation}%
by an angle $\gamma _{c}$ apart from a global phase $\gamma _{c}/2$, where $%
\hat{\sigma}=\left( \sigma _{x},\sigma _{y},\sigma _{z}\right) $ are the
pauli's spin operators. Hence, arbitrary rotation of the qubit can be
accomplished by an adiabatic process.

\begin{figure}[tbp]
\includegraphics[width=6cm]{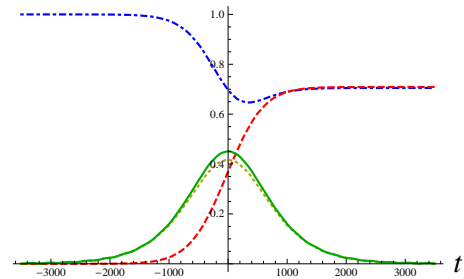}
\caption{(Color online) The norm of amplitudes for state $\left\vert
1\right\rangle $ (blue dot-dashed line), $\left\vert 2\right\rangle $ (red
dashed line), $\left\vert 3\right\rangle $ (yellow dotted line) and $%
\left\vert 4\right\rangle $ (green solid line) as functions of time in the
double-excited model described by $H_{I}$ in Eq.(\protect\ref{eq2:3}). The
horizontal axe is in units of ns. }
\label{fig3:1}
\end{figure}

We now design a rotation of the qubit about $x$ with an angle $\pi /2$ to
support the above analysis. The rotation about axe $x$ is obtained by
choosing fixed angles $\chi =-\pi /4$ and $\phi =0$ in Eq.(\ref{eq2:5}). For
an adiabatic evolution, $\Omega \left( t\right) $ and $\psi \left( t\right) $
are required to be smooth functions which vanish at infinity. We take the
envelope and phase of Rabi frequencies as follow
\begin{subequations}
\label{eq2:11}
\begin{eqnarray}
\Omega \left( t\right) &=&\frac{\sqrt{3}\omega _{0}\sin \eta }{\sqrt{2-\frac{%
3}{2}\sin ^{2}\eta }}, \\
\psi \left( t\right) &=&\frac{\pi }{2}-\arctan \frac{2\cos \eta }{\sin \eta }%
,
\end{eqnarray}%
so that the vector $\left\vert D_{2}\right\rangle $ initial at north point
undergoes an adiabatic evolution along a closed path with a solid angle $%
\gamma _{c}=\pi /2$ in the BS2, where
\end{subequations}
\begin{equation}
\eta =\frac{\pi }{2}\left[ 1+\tanh \left( \alpha t\right) \right] .
\label{eq2:12}
\end{equation}%
Parameter $\alpha $ characterizes the bandwidth of pulses. Having the
designed pulses (\ref{eq2:11}), now we can test our main theoretical results
(\ref{eq2:9}) with the aid of numerical calculations. In order to do the
calculation, we first assume that the system is initially populated in state
$\left\vert 1\right\rangle $. Then we numerical solve the time-dependent Schr%
\"{o}dinger equation described by Hamiltonian in Eq.~(\ref{eq2:3}) with $%
\omega _{0}=20MHz$, $\alpha =1MHz$. In Fig.~\ref{fig3:1}, we plot the norms
of amplitudes for all four states as functions of time.
It can be found that there are no population in the two excited states
(yellow dotted and green solid lines) before and after applying the pulses,
however the amplitude of the initial state $\left\vert 1\right\rangle $
(blue dot-dashed line) has changed from $1$ to $1/\sqrt{2}$, and the
amplitude of the degenerate state $\left\vert 2\right\rangle $ increases
from $0$ to $1/\sqrt{2}$. Therefore, a rotation of the qubit about $x$ with
an angle $\pi /2$ has been obtained. The non-vanishing amplitude of the
double excited states during the adiabatic evolution depicts the
non-adiabatic transition between two dark states.

Notice that the optical transitions in NV centers are original associated to
an orbital-doublet, spin-triplet excited state and the spin-triplet ground
state, the actual operation $U$ is presented by $9\times 9$ matrices. To
check how well our theory works for actualy nine-level systems, we use the
operator fidelity~\cite{WangXG} $F=\left\vert \text{Tr}U_{C}^{\dag
}U\right\vert /2$ as a measure of how accurate the ideal gate $U_{C}$ can be
achieved. Taking the spin-orbit and spin-spin parametersf from Ref.\cite%
{Hollen13}, and letting the Rabi frequency amplitudes, the pulse widths and
the detuning are same as those in the above, we numerical solve the equation
for the unitary evolution operator with the initial condition $U\left(
t_{0},t_{0}\right) =I$. The actual operation $U$ is obtained at the time $%
t_{f}$ that the applied pulses have turned off. Then we obtain the operator
fidelity on the order of $99.9\%$.


\section{\label{Sec:4}Conclusion}

In conclusion, we have shown that it is possible to establish an
excited-double four-level system in the spin-triplet gound state and an
orbital-doublet, spin-triplet excited state of the NV center due to the
spin--orbit and spin--spin interactions, which allow us to apply two pump
and two Stokes laser pulses. It is found that two dark states exist when
both the pump and stokes fields maintain two-photon resonance and the
intensities of two pump (Stokes) fields take the same values. Consequently,
we propose a rotation procedure by encoding the qubit in the ground
degenerate states. We first map the information of the qubit on one dark
state and its corresponding bright state defined by the pump and Stokes
pulses. Then we adiabatically tune the Rabi frequencies and phases of the
pump and Stokes pulses by keeping nonvanishing two-photon resonance so that
the NV center undergoes a cyclic evolution. The adiabatical process first
transfers some components of the bright state to the double excited states,
afterwords, bring these components back to the bright state accompanied by a
phase shift $\gamma _{c}$. Therefore, a rotation of the qubit is acheived
due to the non-Abelian geometric phase produced by the nonadiabtic
transition between the two degenerate dark states. The axis and angle of
rotation is determined by the parameters of the laser fields.

This work is supported by the Program for New Century Excellent Talents in
University (NCET-08-0682), NSFC No.~11074071, and No.~11105050, NFRPC
2012CB922103, PCSIRT No.~IRT0964,
the Key Project of Chinese Ministry of Education (No.~210150), the Research
Fund for the Doctoral Program of Higher Education No. 20104306120003, Hunan
Provincial Natural Science Foundation of China(11JJ7001), and Scientific
Research Fund of Hunan Provincial Education Department (No.~11B076). We
thanks Prof. Renbao Liu and Dr. Nan Zhao for useful discussions.


\begin{thebibliography}{99}
\bibitem{Bpr281} R. Bhandari, Phys. Rep. \textbf{281}, 1 (1997).

\bibitem{Berry84} M. V. Berry, Proc. R. Soc. London, Ser. A \textbf{392}, 45
(1984).

\bibitem{Wilczek} F. Wilczek and A. Zee, Phys. Rev. Lett. \textbf{52}, 2111
(1984).

\bibitem{RMP01523} J. Dalibard, F. Gerbier, G. Juzeli\={u}nas, P. \"{O}berg,
Rev. Mod. Phys. \textbf{83}, 1523 (2011).

\bibitem{sunPRD} Chang-Pu Sun, Mo-Lin Ge, Phys. Rev. D \textbf{41}, 1349
(1990).

\bibitem{graphene} A.H. Castro Neto, F. Guinea, N.M.R. Peres, K.S.
Novoselov, and A.K. Geim, Rev. Mod. Phys. \textbf{81}, 109 (2009).

\bibitem{HQC} P. Zanardi and M. Rasetti, Phys. Lett. A \textbf{264}, 94
(1999); J. Pachos, P. Zanardi and M. Rasetti, Phys. Rev. A \textbf{61},
010305 (1999).

\bibitem{JPCM18} J. Wrachtrup, and F. Jelezko, J. Phys.: Condens. Matter
\textbf{18}, S807 (2006).

\bibitem{Lukins316} M.V.G. Dutt, L. Childress, L. Jiang \textit{et al. }%
Science \textbf{316}, 1312 (2007).

\bibitem{greent06} A.D Greentree, P. Olivero, M. Draganski \textit{et al. }%
J. Phys.: Condens. Matter \textbf{18}, S825 (2006).

\bibitem{stoneham21} A.M. Stoneham, A.H Harker and G.W Morley, J. Phys.:
Condens. Matter \textbf{21}, 364222 (2009).

\bibitem{PNAS107} J. R. Weber, W. F. Koehl, J. B. Varley \textit{et al. }%
PNAS \textbf{107}, 8513 (2010).

\bibitem{Lenef96} A. Lenef and S. C. Rand, Phys. Rev. B \textbf{53}, 13441
(1996).

\bibitem{Hollen13} M W Doherty, N B Manson, P Delaney and L C L Hollenberg,
New J. Phys. \textbf{13}, 025019 (2011).

\bibitem{LukinNJP13} J R Maze, A Gali, E Togan, \textit{et al.} New J. Phys.
\textbf{13}, 025025 (2011).

\bibitem{Jelezko92} F. Jelezko, T. Gaebel, I. Popa, A. Gruber, J. Wrachtrup,
Phys. Rev. Lett. \textbf{92}, 076401 (2004).

\bibitem{Neumann6} P. Neumann, R. Kolesov, B. Naydenov \textit{et al.} Nat.
Phys. \textbf{6}, 249 (2010).

\bibitem{DuPRL105} Fazhan Shi, Xing Rong, Nanyang Xu \textit{et al.} Phys.
Rev. Lett. \textbf{105}, 040504 (2010).

\bibitem{Fuchs11} G. D. Fuchs, G. Burkard, P.V. Klimov and D.D. Awschalom,
Nat. Phys. \textbf{7}, 789 (2011).

\bibitem{Hanson320} R. Hanson, V. V. Dobrovitski, A. E. Feiguin, O. Gywat,
D. D. Awschalom, Science \textbf{320}, 352 (2008);

\bibitem{Zhaonat6} Nan Zhao, Jian-Liang Hu, Sai-Wah Ho, Tsz-Kai Wen, R. B.
Liu, Nat. Nanotech. \textbf{6}, 242 (2011).

\bibitem{Zhao106} Nan Zhao, Zhen-Yu Wang, Ren-Bao Liu, Phys. Rev. Lett.
\textbf{106}, 217205 (2011).

\bibitem{NP5(11)335} K. Dholakia and T. \v{C}i\v{z}m\'{a}r, Nat. Photon.
\textbf{5}, 335 (2011).

\bibitem{NP4(10)211} D.V. Thourhout and J. Roels, Nat. Photon. \textbf{4},
211 (2010).

\bibitem{NVSphotons} C. Kurtsiefer, S. Mayer, P. Zarda, and H. Weinfurter,
Phys. Rev. Lett. \textbf{85}, 290 (2000).

\bibitem{batalov100} A. Batalov, C. Zierl, T. Gaebe \textit{et al.} Phys.
Rev. Lett. \textbf{100}, 077401 (2008).

\bibitem{cptNV97} C. Santori, P. Tamarat, P. Neumann \textit{et al.} Phys.
Rev. Lett. \textbf{97}, 247401 (2006).

\bibitem{Togan466} E. Togan, Y. Chu, A. S. Trifonov \textit{et al. }Nature
466, 730 (2010).

\bibitem{WangXG} Xiaoguang Wang, Zhe Sun, and Z. D. Wang, Phys. Rev. A 79,
012105 (2009).
\end{thebibliography}
\end{document}